\documentclass[useAMS,usenatbib]{mn2e}

\usepackage{amsmath}
\usepackage{graphicx}
\usepackage{wasysym}
\allowdisplaybreaks

\begin{document}

\title[Locations of stationary/periodic solutions]
{Locations of stationary/periodic solutions in mean motion resonances
according to properties of dust grains}
\author[P. P\'{a}stor]{
P.~P\'{a}stor\thanks{pavol.pastor@hvezdarenlevice.sk}\\
Tekov Observatory, Sokolovsk\'{a} 21, 934~01 Levice, Slovak Republic}

\date{}

\pagerange{\pageref{firstpage}--\pageref{lastpage}} \pubyear{2015}

\maketitle

\label{firstpage}

\begin{abstract}
The equations of secular evolution for dust grains in mean
motion resonances with a planet are solved for stationary points.
This is done including both Poynting--Robertson effect and stellar wind.
The solutions are stationary in semimajor axis, eccentricity, and resonant
angle, but allow the pericentre to advance. The semimajor axis of stationary
solutions can be slightly shifted from the exact resonant value.
The periodicity of the stationary solutions in a reference frame
orbiting with the planet is analytically proved. The existence
of periodic solutions in mean motion resonances means that analytical
theory enables for dust particles also infinitely long capture times.
The stationary solutions are periodic motions to which the eccentricity
asymptotically approaches and around which the libration occurs.

Using numerical integration of equation of motion are successfully
found initial conditions corresponding to the stationary solutions.
Numerically and analytically determined shifts of the semimajor
axis form the exact resonance for the stationary solutions
are in excellent agreement.

The stationary solutions can be plotted by locations of pericenters
in the reference frame orbiting with the planet. The pericenters
are distributed in the space according to properties of dust particles.
\end{abstract}


\section{Introduction}
\label{sec:introduction}

A lenticular cloud of dust centered on the Sun with its main axis
lying in the ecliptic plane creates well known zodiacal light.
Discoveries and improving observations of debris disks around main-sequence
stars \citep*[e.g.][]{holland,debes,schneider,lagrange} motivate to
understand better the dynamics of the dust particles in such environments.
Sharp inner edges, clumps and asymmetries observed in the debris disks
are commonly interpreted as signatures of a planet created in earlier
stages of the disk evolution and located near or within the disks
\citep*[e.g.][]{wyatt,KH,SK9,RMH}. Such a planet can gravitationally capture
the dust particles in the so called mean motion resonances (MMRs)
and change the morphology of the disk.

Mean motion resonances occur when a ratio of orbital periods
of the dust particle and the planet is near a ratio of two small
natural numbers. The MMRs in a circular restricted three-body
problem (CRTBP) with a stellar electromagnetic radiation and a stellar
wind include numerous aspects which are directly applicable to astrophysical
systems, ranging from the resonant dust ring close to the Earth's orbit
\citep[e.g.][]{shepherd,IRAS,COBE,reach,points} through dust
in the Edgeworth--Kuiper belt zone \citep[e.g.][]{LZ1999,MM,holmes,KS}
to structures observed in debris disks \citep*[e.g.][]{KH,SK,SMW}.
Analytical results can be applied into semi-analytic models of resonant
dust just to improve computation speed of resonant disk patterns
\citep[e.g.][]{wyatt,KH,SK9,RMH,SMW}.

In this work we show that the CRTBP with radiation comprises also
periodic solutions for the motion of the dust particle in an exterior MMR
(the orbital period of the dust particle is larger than the orbital
period of the planet in the exterior MMRs). The periodic solutions
exist at ``universal eccentricities'' found by \citet{FM2}.
The eccentricity of dust orbit during a capture in the exterior
MMR approaches the universal eccentricity \citep[e.g.][]{LZ1997,AA2,SK9}.
If such periodic solutions exist, then maximal capture times for
the dust particles in the exterior MMRs should be infinite.

Using adiabatic invariant theory \citet{gomes95} found a criterion determining
whether the libration amplitude increases or decreases for dissipative forces
which lead to an evolution of the eccentricity characterised with
asymptotic approach of the eccentricity to the universal eccentricity.
Results of averaged theory in \citet{gomes95} showed that the libration
amplitude must increase for the dust particles in all exterior resonances
in the planar CRTBP with the Poynting--Robertson (PR) effect.
\citet{gomes95} results allow also the existence of the periodic
solutions.

Periodic solutions in 1/1 resonance were already found to exist
\citep*{LZJ,points}. The existence of secularly stationary solutions in
the vicinity of the triangular Lagrangian equilibrium points in
spatial-circular, planar-elliptic and spatial-elliptic restricted three-body
problem with the PR effect was recently confirmed in \citet{triangular}.

The content of this paper is the following.
In Section \ref{sec:resonantequations}, are presented averaged resonant
equations for the MMRs in the planar CRTBP with arbitrary non-gravitational
effects for which secular theory can be applied (general case).
In Section \ref{sec:conditions}, are determined conditions for
the stationary points in the general case.
In Section \ref{sec:periodicity}, is shown that for astrophysical
systems with rotational symmetry an evolution during the stationary
solution in the MMR is always periodic.
In Section \ref{sec:application}, are derived the conditions for
the stationary points when the non-gravitational effects are
the PR effect and the radial stellar wind.
In Section \ref{sec:statio}, we determine the stationary solutions
for the considered non-gravitational effects, we discuss properties
of the solutions and present pericenters of the orbits corresponding
to the stationary solutions in a reference frame orbiting with the Earth.
Conclusions are drawn in Section \ref{sec:conclusions}.

\section{Averaged resonant equations for dust particle influenced
by non-gravitational effects}
\label{sec:resonantequations}

Orbital evolution in a dynamical system is often averaged over a
period of the dynamical system (secular framework). The secular framework
also enables define a stationarity. In the secular framework
the stationarity is obtained if one or more orbital elements
remain constant after the averaging process. In the MMRs the semimajor
axis of the particle orbit is also secularly stationary after the averaging
over an libration period. In what follows the word secularly before
the stationary will be omitted. A dust particle can be captured
in an MMR with a planet in the planar CRTBP and simultaneously be affected
by non-gravitational effects. In the general case secular variations
of a particle orbit, caused by the non-gravitational effects, can depend
on spatial orientation of the orbit. In this case resonant equations
averaged over a synodic period have form \citep{stab}:
\begin{align}\label{transformed}
\frac{dk}{dt} = {} & \frac{\alpha}{L} \frac{\partial R}{\partial h} -
      h \frac{d \sigma_{1}}{dt} -
      \frac{k \alpha}{L ( 1 + \alpha )} \frac{dL}{dt} -
      \frac{k \alpha}{L e^{2}} \left \langle P_{2} \right \rangle -
      h \left \langle Q_{2} \right \rangle ~,
\notag \\
\frac{dh}{dt} = {} & - \frac{\alpha}{L} \frac{\partial R}{\partial k} +
      k \frac{d \sigma_{1}}{dt} -
      \frac{h \alpha}{L ( 1 + \alpha )} \frac{dL}{dt}
\notag \\
& - \frac{h \alpha}{L e^{2}} \left \langle P_{2} \right \rangle +
      k \left \langle Q_{2} \right \rangle ~,
\notag \\
\frac{dL}{dt} = {} & s \left ( h \frac{\partial R}{\partial k} -
      k \frac{\partial R}{\partial h} \right ) +
      \left \langle P_{1} \right \rangle ~,
\notag \\
\frac{d \sigma_{1}}{dt} = {} & \frac{p + q}{q} n_{\text{P}} - s n +
      \frac{2 s a}{L} \frac{\partial R}{\partial^{\star} a}
\notag \\
& - \frac{\alpha s}{L ( 1 + \alpha )}
      \left ( k \frac{\partial R}{\partial k} +
      h \frac{\partial R}{\partial h} \right ) +
      s \left \langle Q_{1} \right \rangle ~.
\end{align}
Here, $k$ $=$ $e \cos \sigma$ and $h$ $=$ $e \sin \sigma$
are the non-canonical resonant variables,
$e$ is the eccentricity of the particle orbit, and
$\sigma$ is the resonant angular variable
$\sigma$ $=$ $\psi / q - \tilde{\omega}$.
The expression for the resonant angular variable contains:
$\psi$ $=$ $( p + q ) \lambda_{\text{P}} - p \lambda$,
two integers $p$ and $q$, the mean longitude $\lambda$,
and the longitude of pericenter $\tilde{\omega}$.
The subscript $\text{P}$ is used for quantities belonging
to the planet. $\alpha$ $=$ $\sqrt{1 - e^{2}}$, $s$ $=$ $p / q$,
and $\sigma_{1}$ $=$ $\psi / q - \tilde{\omega}_{\text{P}}$.
$R$ is the disturbing function of the planar CRTBP.
$L$ $=$ $\sqrt{\mu a}$, where, $\mu$ $=$ $G_{0} M_{\star}$,
$G_{0}$ is the gravitational constant, $M_{\star}$ is the mass
of the star, and $a$ is the semimajor axis of the particle orbit.
$n$ $=$ $\sqrt{\mu / a^{3}}$ is the mean motion of the particle.
$\partial^{\star} a$ denotes the partial derivative with
respect to the semimajor axis calculated with an assumption
that the mean motion of the particle $n$ is not
a function of the semimajor axis \citep[see e.g.][]{danby}.
We use angle $\sigma_{\text{b}}$ defined so that the mean
anomaly $M$ can be computed from $M$ $=$ $n t$ $+$ $\sigma_{\text{b}}$
\citep*{fund}. In this case the following relations hold between
the averaged partial derivatives of the disturbing function \citep{stab}
\begin{align}\label{step}
\frac{\partial R}{\partial \sigma_{\text{b}}} &= - s
\frac{\partial R}{\partial \sigma} ~,
\notag \\
\frac{\partial R}{\partial \tilde{\omega}} &= - \frac{p + q}{q}
\frac{\partial R}{\partial \sigma} ~,
\notag \\
\frac{\partial R}{\partial \sigma} &= k \frac{\partial R}{\partial h} -
h \frac{\partial R}{\partial k} ~,
\notag \\
e \frac{\partial R}{\partial e} &= k \frac{\partial R}{\partial k} +
h \frac{\partial R}{\partial h} ~.
\end{align}
From the definitions after Eqs. (\ref{transformed}) we have
$\sigma$ $=$ $\psi / q - \tilde{\omega}$ and
$\sigma_{1}$ $=$ $\psi / q - \tilde{\omega}_{\text{P}}$.
Therefore, the last equation in Eqs. (\ref{transformed}) reads
\begin{equation}\label{sigmas}
\frac{d \sigma_{1}}{dt} = \frac{d \sigma}{dt} +
\frac{d \tilde{\omega}}{dt} ~,
\end{equation}
where we have used the fact that $\tilde{\omega}_{\text{P}}$ $=$ 0
in the CRTBP. When we again use the definitions after Eqs. (\ref{transformed})
for $L$ $=$ $\sqrt{\mu a}$, $k$ $=$ $e \cos \sigma$ and
$h$ $=$ $e \sin \sigma$ we can obtain, using some analytic
manipulations, the following useful relations
\begin{align}\label{kh}
\frac{da}{dt} &= \frac{2 a}{L} \frac{dL}{dt} ~,
\notag \\
\frac{de}{dt} &= \frac{1}{e} \left ( k \frac{dk}{dt} +
      h \frac{dh}{dt} \right ) ~,
\notag \\
\frac{d \sigma}{dt} &= \frac{1}{e^{2}} \left ( k \frac{dh}{dt} -
      h \frac{dk}{dt} \right ) ~.
\end{align}
Eqs. (\ref{step}), Eq. (\ref{sigmas}) and Eqs. (\ref{kh}) can be used
in order to rewrite Eqs. (\ref{transformed}) as follows
\begin{align}\label{evolution}
\frac{da}{dt} = {} & \frac{2 a}{L}
      \left ( - s \frac{\partial R}{\partial \sigma} +
      \left \langle P_{1} \right \rangle \right ) ~,
\notag \\
\frac{de}{dt} = {} & \frac{\alpha}{L e}
      \left \{ \left [ 1 + s \left ( 1 - \alpha \right ) \right ]
      \frac{\partial R}{\partial \sigma} -
      \left ( 1 - \alpha \right ) \left \langle P_{1} \right \rangle -
      \left \langle P_{2} \right \rangle \right \} ~,
\notag \\
\frac{d \tilde{\omega}}{dt} = {} & \frac{\alpha}{L e}
      \frac{\partial R}{\partial e} - \left \langle Q_{2} \right \rangle ~,
\notag \\
\frac{d \sigma}{dt} = {} & \frac{p + q}{q} n_{\text{P}} - s n -
      \frac{\alpha}{L e} \left [ 1 + s \left ( 1 - \alpha \right ) \right ]
      \frac{\partial R}{\partial e}
\notag \\
& + \frac{2 s a}{L} \frac{\partial R}{\partial^{\star} a} +
      s \left \langle Q_{1} \right \rangle +
      \left \langle Q_{2} \right \rangle ~.
\end{align}
The system given by Eqs. (\ref{evolution}) enables to study
the secular orbital evolution of the dust particle captured in
the specific MMR given by the resonant numbers $p$ and $q$
under the action of the non-gravitational effects. The expression
of Eqs. (\ref{transformed}) using the orbital elements and
the resonant angular variable, given by Eqs. (\ref{evolution}), will be useful
for a determination of properties of the stationary solutions as we will
see later. For the averaged non-conservative terms $\langle Q_{i} \rangle$
and $\langle P_{i} \rangle$, $i$ $=$ 1, 2 hold \citep{stab}
\begin{align}\label{relations}
\left \langle Q_{1} \right \rangle &= -
      \left \langle \frac{d \tilde{\omega}}{dt} \right \rangle_{\text{EF}} -
      \left \langle \frac{d \sigma_{\text{b}}}{dt} +
      t \frac{dn}{dt} \right \rangle_{\text{EF}} ~,
\notag \\
\left \langle Q_{2} \right \rangle &= -
      \left \langle \frac{d \tilde{\omega}}{dt} \right \rangle_{\text{EF}} ~,
\notag \\
\left \langle P_{1} \right \rangle &= \frac{n a}{2}
      \left \langle \frac{da}{dt} \right \rangle_{\text{EF}} ~,
\notag \\
\left \langle P_{2} \right \rangle &= - \frac{n a ( 1 - \alpha )}{2}
      \left \langle \frac{da}{dt} \right \rangle_{\text{EF}} -
      \frac{n a^{2} e}{\alpha}
      \left \langle \frac{de}{dt} \right \rangle_{\text{EF}} ~,
\end{align}
where $\langle da / dt \rangle_{\text{EF}}$,
$\langle de / dt \rangle_{\text{EF}}$,
$\langle d \tilde{\omega} / dt \rangle_{\text{EF}}$, and
$\langle d \sigma_{\text{b}} / dt$ $+$ $t$ $dn / dt \rangle_{\text{EF}}$
are caused by the non-gravitational effects only. The averaged values
on the right-hand sides in Eqs. (\ref{relations}) can be obtained using
the Gaussian perturbation equations of celestial mechanics
\citep[e.g.][]{danby}. Their specific values for the PR
effect and the radial stellar wind will be shown later. The equations
in this section are derived for the general case.

\section{Conditions for stationary points}
\label{sec:conditions}

The stationary points in the $kh$ plane are determined by
\begin{equation}\label{fixed}
\frac{dk}{dt} = \frac{dh}{dt} = 0 ~.
\end{equation}
If Eqs. (\ref{fixed}) hold in a case without the non-gravitational
effects (conservative problem), then they affect also the evolution
of the semimajor axis. This is due to the fact that the condition
$de / dt$ $=$ 0 implies $da / dt$ $=$ 0. This can be most easily seen
if we calculate time derivative of the eccentricity averaged over
the synodic period using the first two equations in Eqs. (\ref{evolution})
and the last two equations in Eqs. (\ref{relations}) (see also Eq. 17 in
\citealt{stab})
\begin{align}\label{eccentricity}
\frac{de}{dt} = {} & \frac{\alpha}{2 a e}
      \left ( \alpha - \frac{p + q}{p} \right ) \frac{da}{dt}
\notag \\
& + \frac{\alpha}{2 a e} \left ( \frac{p + q}{p} - \alpha \right )
      \left \langle \frac{da}{dt} \right \rangle_{\text{EF}} +
      \left \langle \frac{de}{dt} \right \rangle_{\text{EF}} ~.
\end{align}
In the conservative problem Eq. (\ref{eccentricity}) gives for $de / dt$ $=$ 0
that $da / dt$ $=$ 0 or $dL / dt$ $=$ 0 \citep[see also][]{FM2}.
Eq. (\ref{eccentricity}) shows that in the case with the non-gravitational
effects $de / dt$ $=$ 0 is not sufficient for $dL / dt$ $=$ 0, in general.
We are interested in the stationary solutions of the system of equations
given by Eqs. (\ref{transformed}) restricted by the same conditions as in
the conservative problem. If we add the condition $dL / dt$ $=$ 0 to
the conditions in Eqs. (\ref{fixed}), then the semimajor axis will be
also stationary similarly as in the conservative problem at the stationary
points in the $kh$ plane. The condition $dL / dt$ $=$ 0 is necessary
for the periodicity of the stationary solutions as we will see later.
Hence, the stationary points will be determined by the substitution
of the following three equations into Eqs. (\ref{transformed})
\begin{equation}\label{stationary}
\frac{dk}{dt} = \frac{dh}{dt} = \frac{dL}{dt} = 0 ~.
\end{equation}
Eqs. (\ref{stationary}) were used also in \citet{FM1} and \citet{sidnes}.

The substitution of Eqs. (\ref{stationary}) into Eqs. (\ref{transformed})
leads to the conditions
\begin{align}\label{conditions}
\left \langle P_{1} \right \rangle -
s \left \langle P_{2} \right \rangle &= 0 ~,
\notag \\
e \frac{\partial R}{\partial e} = k \frac{\partial R}{\partial k} +
      h \frac{\partial R}{\partial h} &=
      \frac{L e^{2}}{\alpha} \left ( \frac{d \sigma_{1}}{dt} +
      \left \langle Q_{2} \right \rangle \right ) ~,
\notag \\
\frac{\partial R}{\partial \sigma} = k \frac{\partial R}{\partial h} -
h \frac{\partial R}{\partial k} &= \left \langle P_{2} \right \rangle ~.
\end{align}
Eqs. (\ref{stationary}) are equivalent with
\begin{equation}\label{trio}
\frac{da}{dt} = \frac{de}{dt} = \frac{d \sigma}{dt} = 0 ~,
\end{equation}
as can be seen from Eqs. (\ref{kh}). The semimajor axis and the position
of the libration center in the $kh$ plane are stationary in the sought for
stationary points.

\section{Periodicity of evolutions corresponding to stationary solutions
in resonances}
\label{sec:periodicity}

In the exact resonance the averaged mean motion of the particle
and the mean motion of the planet have a ratio that is exactly
equal to the ratio of two natural numbers. The periodicity
of evolutions corresponding to the stationary solutions
in the exact resonances  with fixed longitude
of pericenter was already showed in \citet{stab}. In this
work we take into account the possibility that the longitude of pericenter
can be variable. Influence of the variable longitude of pericenter on
the periodicity of the evolutions during the stationary
solutions will be determined. Beside the variable longitude of pericenter
in this work we consider also non-exact resonances. In this section,
the symbols used denote the un-averaged quantities. We assume that
the physical parameters of the planar CRTBP with the non-gravitational effects
do not change during the synodic period. In our case, the un-averaged
resonant angular variable is
\begin{equation}\label{setup}
\sigma = \frac{p + q}{q} \lambda_{\text{P}} - s \lambda - \tilde{\omega} ~,
\end{equation}
with the mean longitude
\begin{equation}\label{lambda}
\lambda = M + \tilde{\omega} = n t + \sigma_{\text{b}} + \tilde{\omega} ~.
\end{equation}
For the time derivative of the mean longitude in Eqs. (\ref{lambda})
we obtain
\begin{equation}\label{dlambdadt}
\frac{d \lambda}{dt} = n + \frac{d \sigma_{\text{b}}}{dt} + t \frac{dn}{dt} +
\frac{d \tilde{\omega}}{dt} ~.
\end{equation}
Using the previous equation we obtain for the time derivative of $\sigma$
in Eq. (\ref{setup})
\begin{equation}\label{dsigma0dt}
\frac{d \sigma}{dt} = \frac{p + q}{q} n_{\text{P}} - s n -
s \left ( \frac{d \sigma_{\text{b}}}{dt} + t \frac{dn}{dt} +
\frac{d \tilde{\omega}}{dt} \right ) -
\frac{d \tilde{\omega}}{dt} ~.
\end{equation}
Let the motion be stationary according to the conditions in
Eqs. (\ref{trio}). Eq. (\ref{dsigma0dt}) can be averaged over the synodic
period. If we use the last condition in Eqs. (\ref{trio}) for
the stationarity of the resonant angular variable in the result
of averaging, we obtain
\begin{align}\label{passage}
{} & \frac{p + q}{q} n_{\text{P}} - \frac{s}{T_{\text{S}}}
      \int_{0}^{T_{\text{S}}} n ~dt -
      \frac{p + q}{q} \frac{1}{T_{\text{S}}} \int_{0}^{T_{\text{S}}}
      \frac{d \tilde{\omega}}{dt} dt
\notag \\
& - \frac{s}{T_{\text{S}}} \int_{0}^{T_{\text{S}}}
      \left ( \frac{d \sigma_{\text{b}}}{dt} +
      t \frac{dn}{dt} \right ) dt = 0 ~.
\end{align}

In \citet{stab} only the exact resonances were considered.
The condition $n_{\text{P}} ~(p + q) / p$ $=$ $( 1 / T_{\text{S}} )$
$\int_{0}^{T_{\text{S}}}$ $n ~dt$ valid for the exact resonances may not
be valid for the non-exact resonances. Therefore, the first two terms
in Eq. (\ref{passage}) may not cancel each other as in the case
for the exact resonances.

The periodicity of evolution corresponding to the stationary solution
in \citet{stab} was demostrated for the non-gravitational effects which
include an interstellar wind. For the star moving through a cloud
of an interstellar matter the interstellar wind is monodirectional
\citep*[for details see][]{baines}. The monodirectional character
of the interstellar wind enables only the stationary solutions for which
the longitude of pericenter is stationary. The third term in
Eq. (\ref{passage}) is caused by the variability (instationarity)
of the longitude of pericenter and in \citet{stab} was equal to zero.

\begin{figure*}
\begin{center}
\includegraphics[width=0.7608024691358024691358024691358\textwidth]{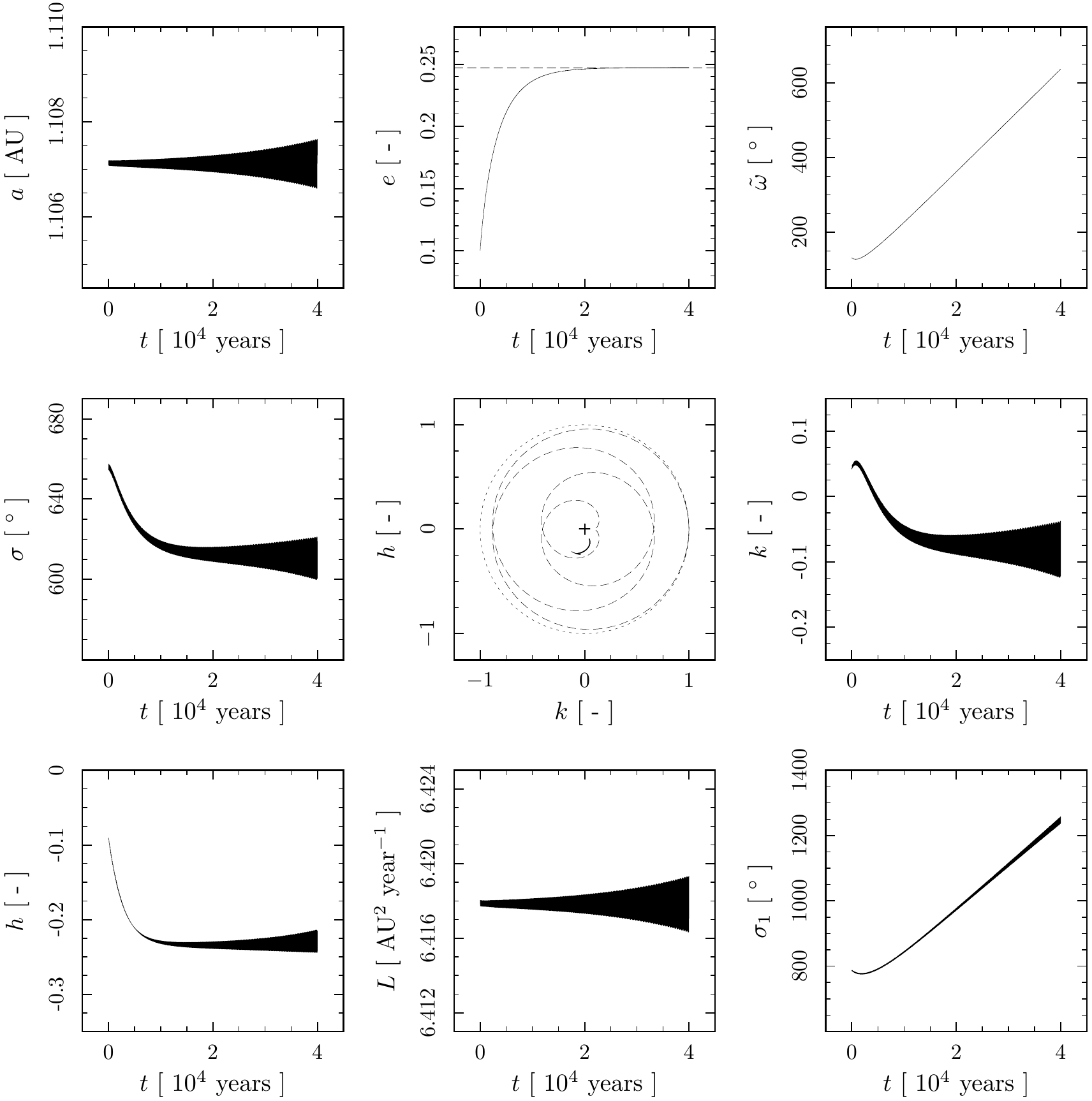}
\end{center}
\caption{Evolution of a dust particle with $R_{\text{d}}$ $=$ 5 $\mu$m,
$\varrho$ $=$ 2 g/cm$^{3}$, and $\bar{Q}'_{\text{pr}}$ $=$ 1 captured
in an exterior mean motion 6/5 orbital resonance in the planar
CRTBP with the PR effect and the radial solar wind which comprises the Earth
and the Sun as the two main bodies. Depicted evolving parameters are
described in the text. The dashed line in the second plot shows
the asymptotic value of the eccentricity. The dashed line in the $kh$
plane indicates the values of $e$ and $\sigma$ that lead to collisions
of the planet with the particle. All evolving parameters are averaged
over the synodic period.}
\label{fig:radiation}
\end{figure*}

\begin{table*}
\caption{The universal eccentricity (the first equation in Eqs. \ref{shift})
for several exterior resonances. The eccentricities implicate from
Eqs. (\ref{stationary}) when the non-gravitational effects are
the PR effect and the radial stellar wind.}
\label{T1}
\begin{tabular}{| c | c | c | c | c | c |}
\hline
\multicolumn{2}{|c|}{1$^{\text{st}}$ order} &
\multicolumn{2}{|c|}{2$^{\text{nd}}$ order} &
\multicolumn{2}{|c|}{3$^{\text{rd}}$ order}\\
\hline
$p / (p + q)$ & universal $e$ & $p / (p + q)$ & universal $e$ &
$p / (p + q)$ & universal $e$\\
\hline
9/8 & 0.1986 & 9/7 & 0.2904 & 8/5 & 0.3972\\
8/7 & 0.2115 & 7/5 & 0.3362 & 7/4 & 0.4331\\
7/6 & 0.2273 & 5/3 & 0.4140 & 5/2 & 0.5505\\
6/5 & 0.2472 & 3/1 & 0.5993 & 4/1 & 0.6654\\
5/4 & 0.2736 & & & &\\
4/3 & 0.3108 & & & &\\
3/2 & 0.3690 & & & &\\
2/1 & 0.4812 & & & &\\
\hline
\end{tabular}
\end{table*}

\begin{figure*}
\begin{center}
\includegraphics[width=0.78395061728395061728395061728395\textwidth]{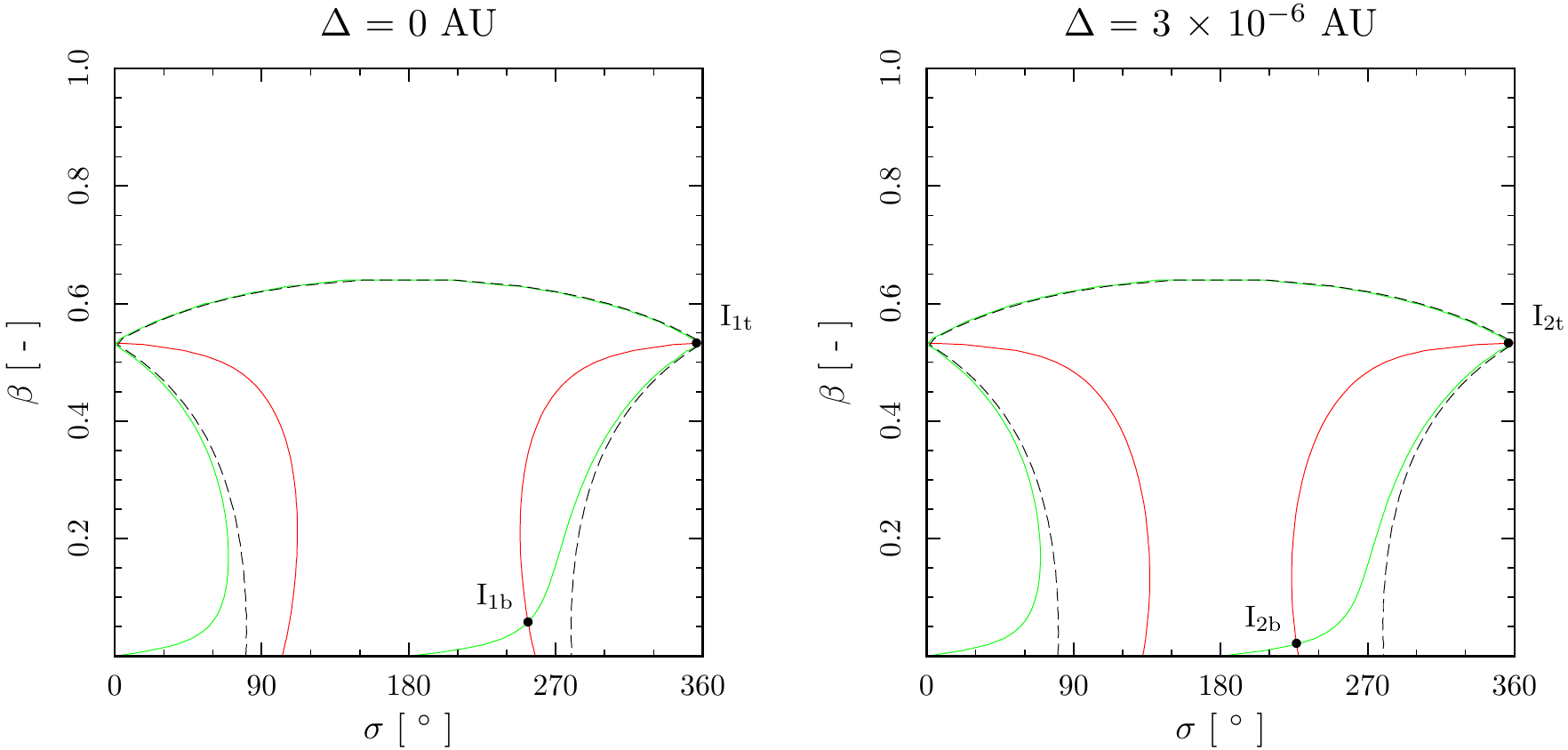}
\end{center}
\caption{Solutions of the conditions for the stationary points in
the planar CRTBP with radiation (Eqs. \ref{shift}). Solutions
of the second equation, determined at the universal eccentricity, are
depicted with red curves. Similarly solutions of the third
equation are depicted with green curves. Dust particles are
captured in an exterior 6/5 MMR with the Earth in a circular
orbit around the Sun. In the left plot is $\Delta$ $=$ 0 used and
the right plot was obtained using $\Delta$ $=$ 3 $\times$ 10$^{-6}$ AU.
Black circles indicate intersections of red and green lines
representing solutions of the system given by Eqs. (\ref{shift}).}
\label{fig:intersections}
\end{figure*}

\begin{figure*}
\begin{center}
\includegraphics[width=0.82253086419753086419753086419753\textwidth]{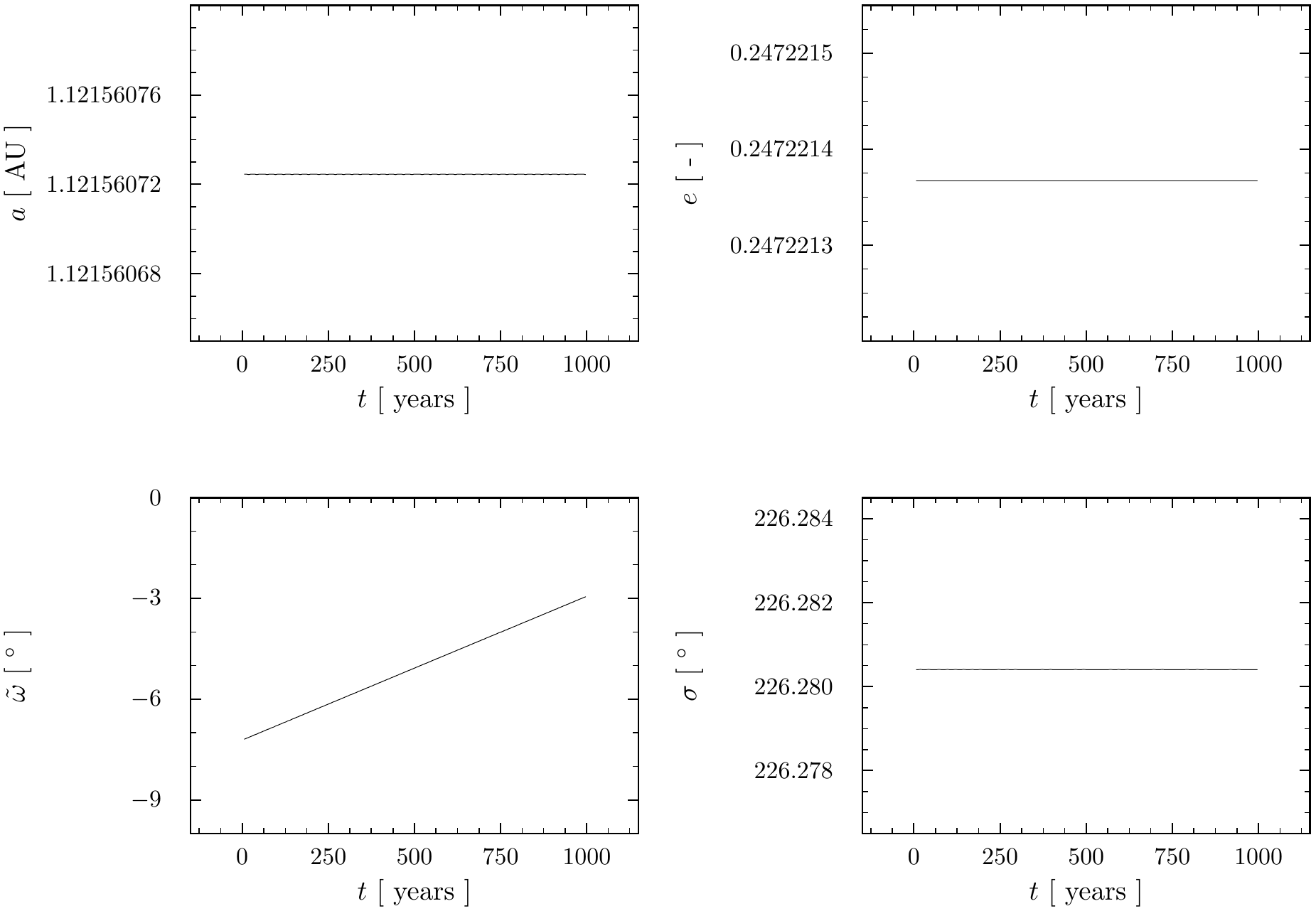}
\end{center}
\caption{Evolution of orbital parameters averaged over the synodic
period obtained from numerical integration of the equation of motion
(Eq. \ref{EOM}) close to the stationary point $\text{I}_{2\text{b}}$ depicted
in Fig. \ref{fig:intersections}. A dust particle with $\beta$ $=$ 0.02027611
and $\bar{Q}'_{\text{pr}}$ $=$ 1 is captured in an exterior mean
motion 6/5 orbital resonance with the Earth in a circular orbit around
the Sun under the action of the PR effect and the radial solar wind.
The stationary solution occurs in a non-exact MMR. Un-averaged evolution
during the first 30 years can be seen in Fig. \ref{fig:evolution}.}
\label{fig:stationarity}
\end{figure*}

\begin{figure*}
\begin{center}
\includegraphics[width=0.80246913580246913580246913580247\textwidth]{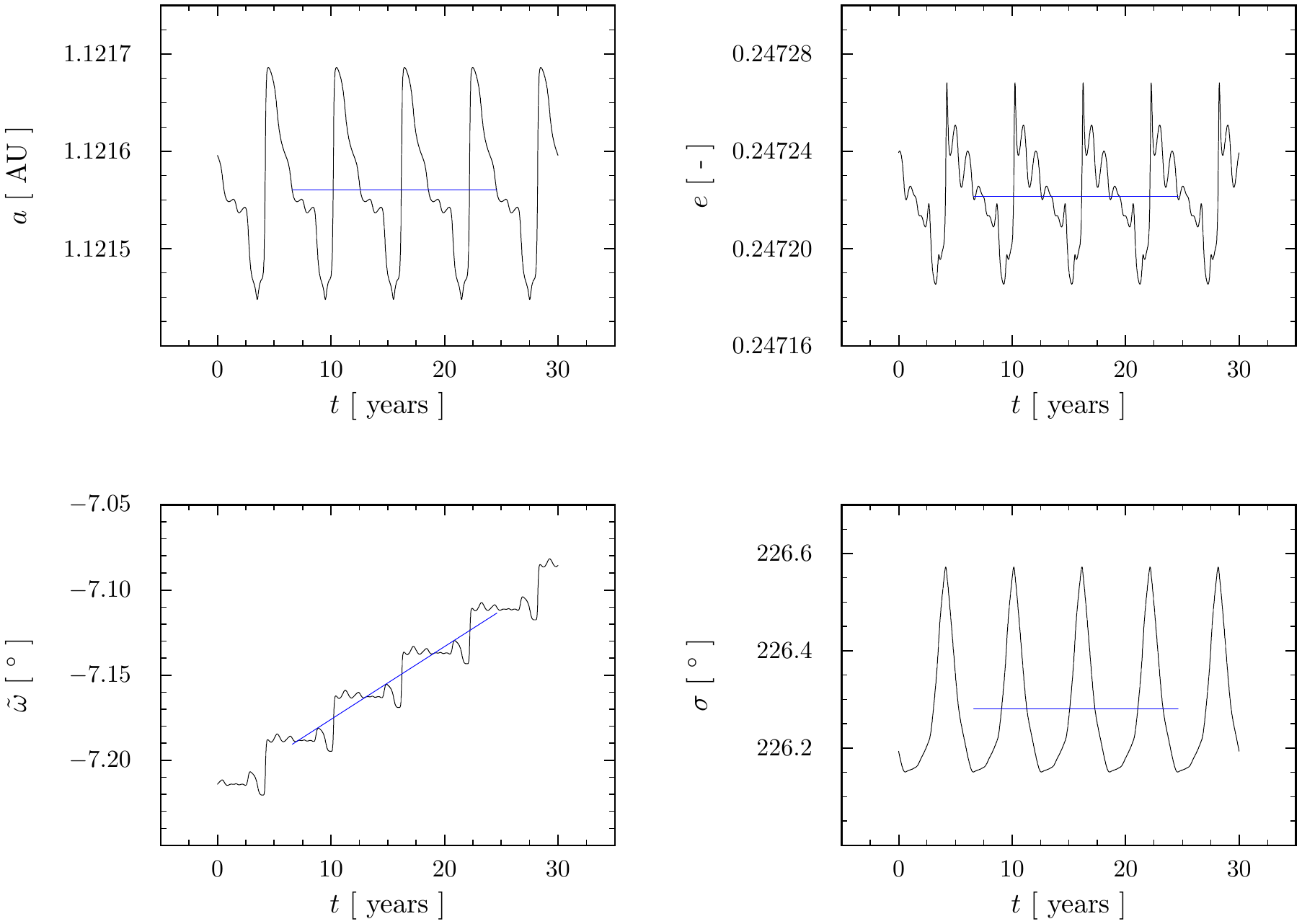}
\end{center}
\caption{Evolution of osculating orbital parameters during initial 30
years of the evolution depicted in Fig. \ref{fig:stationarity} (black line).
For the sake of comparison also the evolution of orbital parameters averaged
over the synodic period is shown (blue line). The averaged evolution is
created only by four points placed in the middle of a given averaged
synodic period. Scales on vertical axes are different from the scales
in Fig. \ref{fig:stationarity} in order to show full evolution
of osculating $a$, $e$ and $\sigma$ during the synodic period.}
\label{fig:evolution}
\end{figure*}

\begin{figure*}
\begin{center}
\includegraphics[width=0.77006172839506172839506172839506\textwidth]{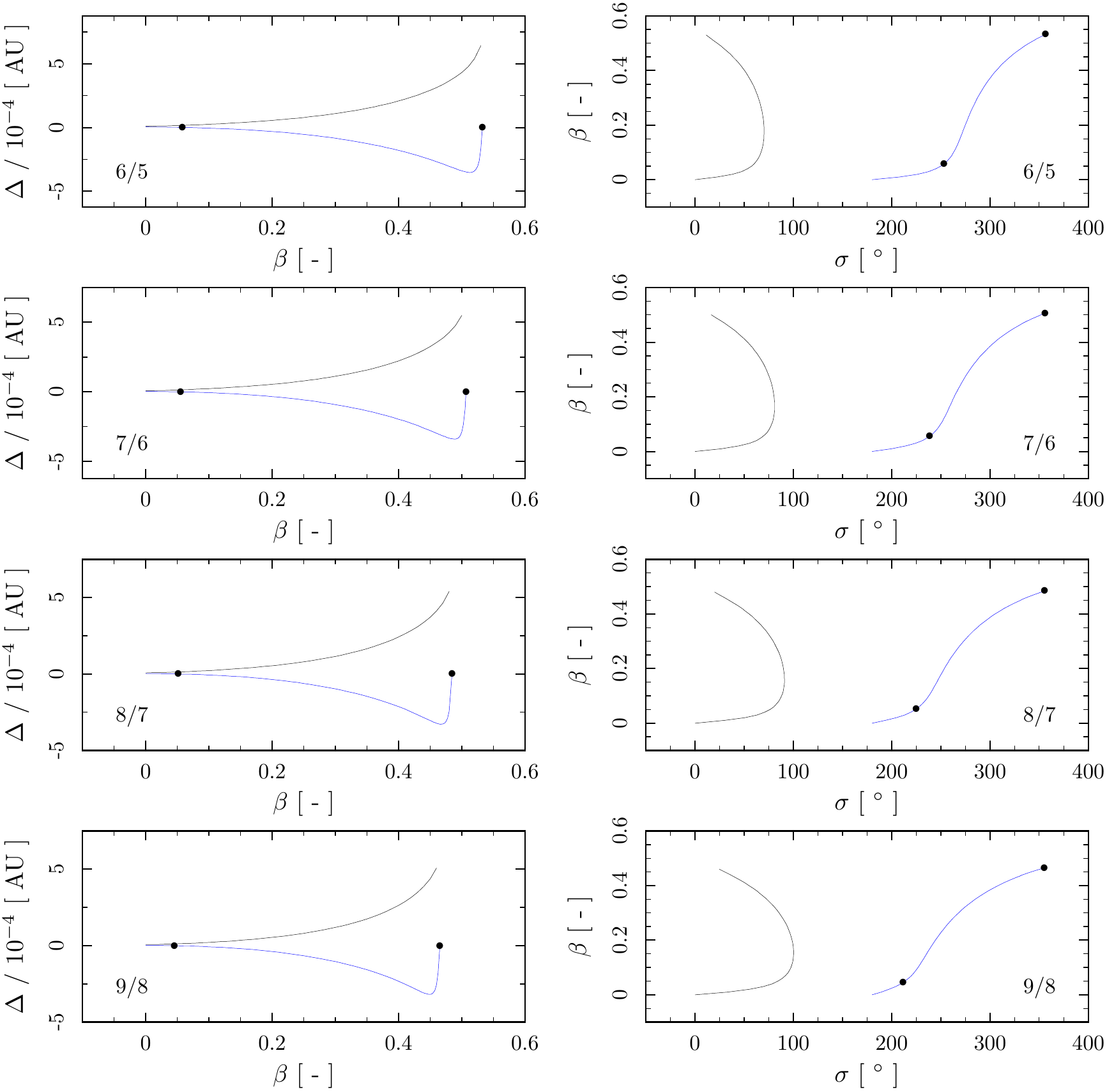}
\end{center}
\caption{Stationary solutions found from
Eqs. (\ref{shift}). Dependencies of the shift from the exact resonance
$\Delta$ on $\beta$ (left column) and dependencies of $\beta$ on
the resonant angular variable $\sigma$ (right column) are shown
for 6/5, 7/6, 8/7 and 9/8 exterior MMRs in a circular Sun-Earth-dust
problem with the PR effect and the radial solar wind. Equal color in
a given row for a given resonance corresponds to the same sets
of the stationary solutions. The semimajor axis of the solutions
can be computed as $a$ $=$ $a_{\text{r}}$ $+$ $\Delta$.
The stationary solutions in the exact resonances are marked with
black circles. The stationary eccentricity is equal to the universal
eccentricity.}
\label{fig:statio}
\end{figure*}

\begin{figure*}
\begin{center}
\includegraphics[width=0.62037037037037037037037037037037\textwidth]{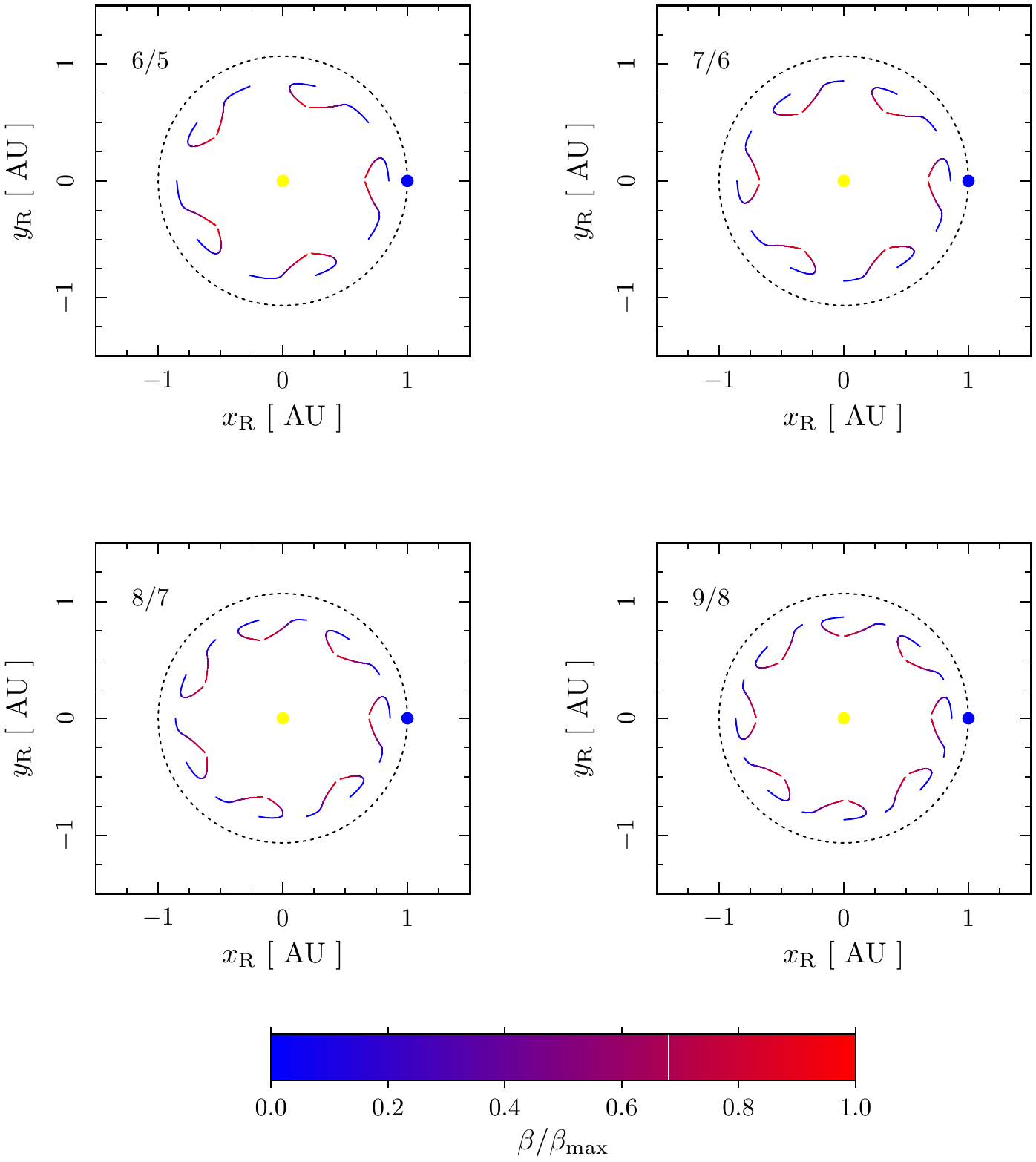}
\end{center}
\caption{Pericenters of orbits corresponding to the stationary solutions
shown in the reference frame orbiting with the Earth for dust particles
captured in 6/5, 7/6, 8/7 and 9/8 exterior mean motion orbital resonances
in a circular Sun-Earth-dust problem with radiation. The eccentricity
of orbits is equal to the universal eccentricity. Intervals for
$\tilde{\omega}$ are calculated from $\sigma$ depicted in the right
column of Fig. \ref{fig:statio} (see text). $\beta_{\text{max}}$
is maximal $\beta$ at which the stationary solution exists and
$\beta_{\text{max}}$ belongs to the 6/5 resonance ($\beta_{\text{max}}$
$=$ 0.532394).}
\label{fig:branches}
\end{figure*}

Eq. (\ref{dlambdadt}) can also be averaged over the synodic period.
If we substitute Eq. (\ref{passage}) in the averaged
Eq. (\ref{dlambdadt}), then we obtain
\begin{equation}\label{average}
\frac{1}{T_{\text{S}}} \int_{0}^{T_{\text{S}}} \frac{d \lambda}{dt} dt =
\frac{p+q}{p} n_{\text{P}} - \frac{q}{p} \frac{1}{T_{\text{S}}}
\int_{0}^{T_{\text{S}}} \frac{d \tilde{\omega}}{dt} dt ~.
\end{equation}
In the equation above we can see that the difference between mean longitudes
after the synodic period $T_{\text{S}}$ depends on the synodic period
and on the averaged time derivative of the longitude of pericenter.
In what follows we determine a relation between these two variables.
Eqs. (\ref{transformed}) were obtained by the averaging over
the synodic period using a cyclic angular variable $\sigma_{2}$
as the integration variable. $\sigma_{2}$ is defined
\citep[see e.g.][]{sidnes} as follows
\begin{equation}\label{cyclic}
\sigma_{2} = \frac{\lambda - \lambda_{\text{P}}}{q} ~.
\end{equation}
The difference between $\sigma_{2}$ at time zero and $\sigma_{2}$
after one synodic period $T_{\text{S}}$ is equal to $2 \pi$.
Implicit differentiation of $\sigma_{2}$ defined in Eq. (\ref{cyclic}) yields
\begin{equation}\label{differentiation}
d \sigma_{2} = \frac{1}{q} \left ( n ~dt + t ~dn +
d \sigma_{\text{b}} + d \tilde{\omega} - n_{\text{P}} ~dt \right ) ~.
\end{equation}
The last equation can be integrated over the synodic period
\begin{align}\label{integration}
2 \pi = {} & \frac{1}{q} \biggl [ \int_{0}^{T_{\text{S}}} n ~dt +
      \int_{0}^{T_{\text{S}}} \left ( \frac{d \sigma_{\text{b}}}{dt} +
      t ~\frac{dn}{dt} \right ) dt
\notag \\
& + \int_{0}^{T_{\text{S}}} \frac{d \tilde{\omega}}{dt} dt -
      n_{\text{P}} ~T_{\text{S}} \biggr ] ~.
\end{align}
Substitution Eq. (\ref{passage}) into Eq. (\ref{integration}) yields
the sought for relation between the synodic period and the averaged
time derivative of the longitude of pericenter
\begin{equation}\label{one}
2 \pi p = n_{\text{P}} T_{\text{S}} -
\int_{0}^{T_{\text{S}}} \frac{d \tilde{\omega}}{dt} dt ~.
\end{equation}
The difference between mean anomalies after one synodic period can be
obtained using
\begin{equation}\label{identity}
\frac{1}{T_{\text{S}}} \int_{0}^{T_{\text{S}}} \frac{d \lambda}{dt} dt =
\frac{1}{T_{\text{S}}} \int_{0}^{T_{\text{S}}} \frac{dM}{dt} dt +
\frac{1}{T_{\text{S}}} \int_{0}^{T_{\text{S}}} \frac{d \tilde{\omega}}{dt} dt
\end{equation}
and Eq. (\ref{one}) in Eq. (\ref{average}). We obtain
\begin{equation}\label{result}
\frac{1}{T_{\text{S}}} \int_{0}^{T_{\text{S}}} \frac{dM}{dt} dt =
\left ( p + q \right ) \frac{2 \pi}{T_{\text{S}}} ~.
\end{equation}
The last equation finally implies
\begin{equation}\label{whole}
M(T_{\text{S}}) - M(0) = \left ( p + q \right ) 2 \pi ~.
\end{equation}
Now, we rewrite Eq. (\ref{one}) as follows
\begin{equation}\label{difference}
\tilde{\omega}(0) - \lambda_{\text{P}}(0) =
\tilde{\omega}(T_{\text{S}}) - \lambda_{\text{P}}(T_{\text{S}}) + 2 \pi p ~,
\end{equation}
where we have used the identity $\lambda_{\text{P}}(T_{\text{S}})$ $-$
$\lambda_{\text{P}}(0)$ $=$ $n_{\text{P}} T_{\text{S}}$.
Eq. (\ref{difference}) shows that a difference in the angle between
the planet and the pericenter of the particle orbit after the synodic
period is an integer multiplied by $2 \pi$. Eq. (\ref{difference}) takes
into account the arbitrary variability of the longitude of pericenter.
Eq. (\ref{whole}) gives that after the synodic period the particle is
in the same position on the orbit as in the time zero. The first two
equations in Eqs. (\ref{trio}) ensure that after the synodic period
the semimajor axis and the eccentricity are the same as in the time zero.
On the basis of Eqs. (\ref{trio}), Eq. (\ref{whole}) and
Eq. (\ref{difference}) we can draw the following conclusions
for the evolution during the stationary solution in
the resonance. If the secular time derivatives of the orbital
elements caused by the non-gravitational effects (non-gravitational
secular variations) do not depend on the longitude
of pericenter, then the evolution will be repeated after the synodic
period (the motion of the dust particle is periodic). The non-gravitational
secular variations do not depend on the longitude of pericenter in
astrophysical systems with rotational symmetry (Appendix \ref{sec:longitude}).
In the cases when the non-gravitational secular variations depend on
the longitude of pericenter the variation of the longitude of pericenter
after one synodic period can cancel the periodicity.

\section{Conditions for stationary points with PR effect
and radial stellar wind}
\label{sec:application}

Let us consider a star which produces an electromagnetic radiation
and a radial stellar wind with one planet in a circular orbit.
In astrophysical modelling of the orbital evolution of dust grains
it is often assumed that a spherical particle can be used as
an approximation to real, the arbitrarily shaped, particle.
The force associated with the action of the electromagnetic
radiation on the moving the spherical dust particle is the PR effect
\citep{poynting,robertson,klacka2004,icarus}. The stellar wind also
affects the dynamics of the dust particles. A relativistically covariant
equation of motion of the dust particle caused by a parallel
beam of corpuscules impinging on the dust particle was derived in
\citet{covsw}. On the basis of the mentioned papers we obtain
the following equation of motion for the spherical dust particle
in a reference frame associated with the star in accuracy to first
order in $v / c$ ($v$ is the speed of the dust particle with respect to
the star and $c$ is the speed of light in vacuum), first order in $u / c$
($u$ is the speed of the stellar wind with respect to the star)
and first order in $v / u$
\begin{align}\label{EOM}
\frac{d \vec{v}}{dt} = {} & - \frac{\mu}{r^{2}}
      \left ( 1 - \beta \right ) \vec{e}_{\text{R}} -
      \frac{G_{0} M_{\text{P}}}{\vert \vec{r} - \vec{r}_{\text{P}} \vert^{3}}
      \left ( \vec{r} - \vec{r}_{\text{P}} \right ) -
      \frac{G_{0} M_{\text{P}}}{r_{\text{P}}^{3}}
      \vec{r}_{\text{P}}
\notag \\
& - \beta \frac{\mu}{r^{2}}
      \left ( 1 + \frac{\eta}{\bar{Q}'_{\text{pr}}} \right )
      \left ( \frac{\vec{v} \cdot \vec{e}_{\text{R}}}{c}
      \vec{e}_{\text{R}} + \frac{\vec{v}}{c} \right ) ~,
\end{align}
where $r$ is the radial distance between the star and the dust particle,
$\vec{e}_{\text{R}}$ is the unit vector directed from the star
to the particle, $\vec{v}$ the velocity of the particle with respect
to the star, $M_{\text{P}}$ is the mass of the planet, $\vec{r}_{\text{P}}$ is
the position vector of the planet with respect to the star, and
$r_{\text{P}}$ $=$ $\vert \vec{r}_{\text{P}} \vert$.
The parameter $\beta$ is defined as the ratio between
the electromagnetic radiation pressure force and the gravitational
force between the star and the particle at rest with
respect to the star. For $\beta$ we have from its definition
\begin{equation}\label{beta}
\beta = \frac{3 L_{\star} \bar{Q}'_{\text{pr}}}{16 \pi c \mu R_{\text{d}}
\varrho} ~,
\end{equation}
where $L_{\star}$ is the stellar luminosity, $\bar{Q}'_{\text{pr}}$ is
the dimensionless efficiency factor for the radiation pressure averaged
over the stellar spectrum and calculated for the radial direction
($\bar{Q}'_{\text{pr}}$ $=$ 1 for a perfectly absorbing sphere),
$c$ is the speed of light in vacuum, and $R_{\text{d}}$ is
the radius of the dust particle with mass density $\varrho$.
$\eta$ is to the given accuracy the ratio of the stellar wind energy to
the stellar electromagnetic radiation energy, both radiated per unit time
\begin{equation}\label{eta}
\eta = \frac{4 \pi r^{2} u}{L_{\star}}
\sum_{i = 1}^{N} n_{\text{sw}~i} m_{\text{sw}~i} c^{2} ~,
\end{equation}
where $m_{\text{sw}~i}$ and $n_{\text{sw}~i}$, $i$ $=$ 1 to $N$, are
the masses and concentrations of the stellar wind particles at a distance
$r$ from the star ($u$ $=$ 450 km/s and $\eta$ $=$ 0.38 for the Sun,
\citealt{covsw}).

The orbital evolution will be given by the evolution of osculating
orbital elements calculated for the case when a central acceleration is
defined by the first Keplerian term in Eq. (\ref{EOM}), namely
$- \mu \left ( 1 - \beta \right ) \vec{e}_{\text{R}} / r^{2}$.
The expressions in the previous sections which contain $L$ and $n$ must
be modified for this central acceleration using a reduced mass
of the star $M_{\star} ( 1 - \beta )$. The modifying equations
are $L$ $=$ $\sqrt{\mu ( 1 - \beta ) a}$ and
$n$ $=$ $\sqrt{\mu ( 1 - \beta ) / a^{3}}$. Using the last term
in Eq. (\ref{EOM}) as a perturbation of the orbital motion we obtain
from the Gaussian perturbation equations of celestial mechanics
the following averaged values in Eqs. (\ref{relations})
\begin{align}\label{secular}
\left \langle \frac{da}{dt} \right \rangle_{\text{EF}} = {} & -
      \frac{\beta \mu}{c a \alpha^{3}}
      \left ( 1 + \frac{\eta}{\bar{Q}'_{\text{pr}}} \right )
      \left ( 2 + 3 e^{2} \right ) ~,
\notag \\
\left \langle \frac{de}{dt} \right \rangle_{\text{EF}} = {} & -
      \frac{\beta \mu}{2 c a^{2} \alpha}
      \left ( 1 + \frac{\eta}{\bar{Q}'_{\text{pr}}} \right ) 5 e ~,
\notag \\
\left \langle \frac{d \tilde{\omega}}{dt} \right \rangle_{\text{EF}} = {} &
      0 ~,
\notag \\
\left \langle \frac{d \sigma_{\text{b}}}{dt} +
t \frac{dn}{dt} \right \rangle_{\text{EF}} = {} & 0 ~.
\end{align}
The substitution of Eqs. (\ref{secular}) into Eqs. (\ref{relations}) gives
\begin{align}\label{PQ_averaged}
\left \langle Q_{1} \right \rangle &= 0 ~,
\notag \\
\left \langle Q_{2} \right \rangle &= 0 ~,
\notag \\
\left \langle P_{1} \right \rangle &= - \frac{\beta \mu n}{2 c \alpha^{3}}
      \left ( 1 + \frac{\eta}{\bar{Q}'_{\text{pr}}} \right )
      \left ( 3 e^{2} + 2 \right ) ~,
\notag \\
\left \langle P_{2} \right \rangle &= -
      \left \langle P_{1} \right \rangle - \frac{\beta \mu n}{c}
      \left ( 1 + \frac{\eta}{\bar{Q}'_{\text{pr}}} \right ) ~.
\end{align}

Now, we return to the conditions for the stationary points during
a capture in the MMRs occurring in the planar CRTBP with non-gravitational
effects given by Eqs. (\ref{conditions}). The substitution
of Eqs. (\ref{PQ_averaged}) in Eqs. (\ref{conditions}) yields for
the PR effect and the radial solar wind
\begin{align}\label{shift}
1 - \frac{3 e^{2} + 2}{2 \alpha^{3}} \frac{p + q}{p} &= 0 ~,
\notag \\
\frac{\alpha}{L e} \left [ 1 + s \left ( 1 - \alpha \right ) \right ]
      \frac{\partial R}{\partial e} &=
      \frac{p + q}{q} n_{\text{P}} - s n +
      \frac{2 s a}{L} \frac{\partial R}{\partial^{\star} a} ~,
\notag \\
\frac{\partial R}{\partial \sigma} &= - \frac{q}{p + q}
      \frac{\beta \mu n}{c}
      \left ( 1 + \frac{\eta}{\bar{Q}'_{\text{pr}}} \right ) ~.
\end{align}
For an interior resonance the first equation in Eqs. (\ref{shift}) does not
have any solution. $p + q$ $<$ $p$ holds in an exterior resonance and
the solution of the first equation in Eqs. (\ref{shift}) is the ``universal
eccentricity'' \citep{FM2,LZ1997}. The universal eccentricities for several
exterior resonances are in Table \ref{T1}.

An example of the evolution approaching to the universal eccentricity
is depicted in Fig. \ref{fig:radiation}. The evolution is obtained from
numerical solution of the equation of motion (Eq. \ref{EOM}) for a dust
particle with $R_{\text{d}}$ $=$ 5 $\mu$m, $\varrho$ $=$ 2 g/cm$^{3}$, and
$\bar{Q}'_{\text{pr}}$ $=$ 1 captured in an exterior mean motion 6/5
orbital resonance with the Earth in a circular orbit around the Sun
under the action the PR effect and the radial solar wind.
The depicted parameters are from left to right and from top to bottom,
as follows: $a$ (semimajor axis), $e$ (eccentricity), $\tilde{\omega}$
(longitude of perihelion), $\sigma$ (resonant angular variable),
$kh$ point, $k$ $=$ $e \cos \sigma$, $h$ $=$ $e \sin \sigma$,
$L$ $=$ $\sqrt{\mu ( 1 - \beta ) a}$, and
$\sigma_{1}$ $=$ $( p + q ) \lambda_{\text{P}} / q$ $-$ $p \lambda / q$.
Collisions of the particle with the planet occur on dashed lines
in the $kh$ plane. The dashed lines cannot be crossed by
the $kh$ point during the evolution in an MMR.

\section{Stationary/periodic solutions}
\label{sec:statio}

Stationary solutions can be found as intersections of solutions
of the second and the third equation in Eqs. (\ref{shift}) at the universal
eccentricity for one chosen semimajor axis and for one chosen longitude
of pericenter. The semimajor axis will be given by a shift $\Delta$
from an exact resonant semimajor axis determined by the condition
$a_{\text{r}}$ $=$ $a_{\text{P}}$ $( 1 - \beta )^{1/3}$
$[ M_{\star} / ( M_{\star} + M_{\text{P}}) ]^{1/3}$
$[ p / ( p + q ) ]^{2/3}$. From this definition we have
$\Delta$ $=$ $a$ $-$ $a_{\text{r}}$. Solutions of the second
equation (red line) and third equation (green line) in Eqs. (\ref{shift})
are depicted in Fig. \ref{fig:intersections}. Collisions of the particle with
the planet occur on curves depicted with dashed lines. Solved system comprises
the Sun and the Earth as two main bodies in the planar CRTBP with radiation.
The equations were solved for all dust particles with $\bar{Q}'_{\text{pr}}$
$=$ 1 which can be in bound orbits around the Sun ($\beta$ $<$ 1) and
are captured in the exterior mean motion 6/5 orbital resonance with
the Earth. The eccentricity of particle orbit is equal to the universal
eccentricity and the longitude of pericenter of particle orbit is equal to
zero for both panels. In reality, the positions of the lines in
Fig. \ref{fig:intersections} should not be depended on the longitude
of pericenter (see Section \ref{sec:periodicity}). Depending on used
numerical averaging method their positions can be found to be slightly
depended on $\tilde{\omega}$. The left panel has $\Delta$ $=$ 0 and
the right panel has $\Delta$ $=$ 3 $\times$ 10$^{-6}$ AU. As can be
seen in Fig. \ref{fig:intersections} the positions of the green lines
are not significantly dependent on $\Delta$ while the positions
of the red lines are. Top boundaries for the collisions in
Fig. \ref{fig:intersections} are obtained when an aphelion
of the particle orbit is located on the planetary orbit. The value
of $\beta_{\text{top}}$ on the boundaries can be calculated as
\begin{equation}\label{betamax}
\beta_{\text{top}} = 1 - \frac{M_{\star} + M_{\text{P}}}{M_{\star}}
\left ( \frac{p + q}{p} \right )^{2}
\left ( \frac{1}{1 + e_{\text{lim}}} -
\frac{\Delta}{a_{\text{P}}} \right )^{3} ~,
\end{equation}
where $e_{\text{lim}}$ is the universal eccentricity. Solutions
of the third equation (green lines) for large values of $\beta$ are close to
the collisions. At large $\beta$ strong gravitational pull from the planet
is needed in order to compensate the influence caused by the PR effect
and the solar wind (Eqs. \ref{secular}). Very large values
of $\beta$ are shown in Fig. \ref{fig:intersections} only for completeness
of the depicted solutions since a capture of the dust particle in an MMR
at these values of $\beta$ is complicated by strong perturbation from
the non-gravitational effects. The stationary solutions are located
in intersections of the red lines and the green lines in Fig.
\ref{fig:intersections}. The intersections are marked with black circles.
Two intersections exist for each panel. The top intersections (subscript t)
are located close to collisions while the bottom intersections are not.

Orbital evolution obtained from numerical solution of the equation
of motion (Eq. \ref{EOM}) during a stationary solution is presented
in Fig. \ref{fig:stationarity}. All evolving parameter are obtained from
averaging over the synodic period (evolution of osculating orbital
parameters during initial 30 years of this numerical solution can be
seen in Fig. \ref{fig:evolution}). The stationary evolution is found
for the stationary point $\text{I}_{2\text{b}}$ depicted in
Fig. \ref{fig:intersections}. Numerically found $\Delta$
of the averaged semimajor axis is 3.0057 $\times$ 10$^{-6}$ AU. This is
in an excellent agreement with value 3 $\times$ 10$^{-6}$ AU predicted
by presented analytical theory for this dust particle (see
Fig. \ref{fig:intersections}). During the time of integration
the longitude of pericenter increased by about 4.2419$^{\circ}$.
At greater resolution of vertical axes in Fig. \ref{fig:stationarity}
it is possible to see oscillations in the evolutions of $a$, $e$ and
$\sigma$. This is caused by finite possibilities of the numerical integration.
The stationary $e$ and $\sigma$ determined by the numerical solution
of the equation of motion are found to be very sensitive on a precision
of the numerical integration. Differences between the values
of the stationary $e$ and $\sigma$ obtained from the presented
analytical theory and the values obtained from the numerical integration
can be caused also by limits of averaging used in the determination
of the analytical values.

We searched the stationary solutions for exterior resonances
6/5, 7/6, 8/7 and 9/8 in a circular Sun-Earth-dust problem with
the PR effect and the radial solar wind. We have chosen these exterior
resonances since the orbits corresponding to the stationary
solutions for exterior resonances of the first order ($q$ $=$ $-$ 1) with
$p$ $<$ 6 reach close to or behind the orbit of Mars for particles
with small $\beta$. We must note that the orbits corresponding to
the stationary solutions for the considered exterior resonances
at large $\beta$ ($\sim$ 0.5) can reach the orbit of Venus.
The stationary solutions are shown in Fig. \ref{fig:statio}.
Dependencies of $\Delta$ on $\beta$ are shown in the left
column of the figure and dependencies of $\beta$ on $\sigma$
are shown in the right column of the figure. Equal color
on both sides corresponds to equal set of the solutions.
We will call these sets of the solutions branches. The bottom
branches of the left-hand side plots begins at $\beta$ $=$ 0
and ends with the solution with $\Delta$ $=$ 0. The stationary
solutions for the exact resonances are marked with black circles.
The top branches of the left-hand side plots ends with
minimal $\beta$ for which a crossing of collision curves
occur if are all stationary $\Delta$ considered. Such a crossing
is located in Fig. \ref{fig:intersections} close to the left and
the right margins of both plots. The crossing of collision lines
corresponds to such a particle on such an orbit that collision
occurs during one orbit before and also after the pericenter
of the particle orbit (in Keplerian approximation of the motion).

According to the results in Section \ref{sec:periodicity} the angle
between the planet and the pericenter of the particle orbit does not change
after the synodic period. In Fig. \ref{fig:branches} are shown pericenters
of the stationary solutions in a reference frame orbiting with
the Earth. The number of stationary branches in this reference frame is
multiplied by $p$ $+$ $q$. This is caused by the periodicity of $\sigma$
with the period $2 \pi$ and by the identity $\tilde{\omega}$ $=$
$-$ $q$ / $( p + q )$ $\sigma$ holding for the particle in pericenter
and the planet located on the $x$-axis (Eq. \ref{setup}). In
the Fig. \ref{fig:branches} we can see that the pericentes
create a ring inside of Earth's orbit. When the particle is
in pericenter, then a difference between angular velocities
of the planet and the particle is minimal. The difference can be also
negative for the considered exterior mean motion resonances. Therefore
the particle can spend longer time in these places with respect
to the planet \citep{KH}.

The stationary solutions for the considered resonances located
on the bottom branches in the left plots in Fig. \ref{fig:statio}
can contribute to a cloud trailing the Earth observed by satellites
{\it IRAS} \citep{IRAS} and {\it COBE} \citep{COBE}. These branches
are located immediately below the Earth in Fig. \ref{fig:branches}.
The contribution from these resonances should lead to a trailing cloud
which should occupy large interval of heliocentric distances. This
is inconsistent with observed shape of the cloud (see Fig. 2b
in \citealt{COBE}). Narrow circumsolar dust ring was observed also
close to the orbit of Venus by the {\it STEREO} mission
\citep{venus}. A special case mean motion 1/1 orbital resonance
produces narrow circumstellar rings since the universal
eccentricity for this resonance is zero. The particles captured in
the 1/1 resonance can stay close to the same place with respect
to the planet and produce the trailing cloud. However there is problem
with capture the dust particles into the 1/1 resonance. It is practically
impossible for a dust particle to drift towards and get trapped in
the 1/1 resonance when the dust particle is approaching the Sun under
the PR effect and radial stellar wind \citep{LZJ}.
The capture into exterior resonances is possible.

\section{Conclusions}
\label{sec:conclusions}

We have investigated stationary solutions of dust orbits for cosmic dust
particles under the action of the PR effect and the radial stellar wind
captured in MMRs with a planet in a circular orbit. The secular resonant
evolution can be described using averaged resonant equations. The averaged
resonant equations yield a system of equations valid for stationary points.

The stationarity of the semimajor axis and of the libration center
($kh$ point) is most often considered \citep{FM1,FM2,sidnes}.
The validity of the conditions for the stationary point in
astrophysical systems with rotational symmetry yields
a periodic motion in the reference frame orbiting
with the planet around the star.

We successfully solved the averaged resonant equations restricted
by the conditions for the stationary points when the non-gravitational
effects are the PR effect and the radial stellar wind. Search
of the stationary points using numerical solution of the equation
of motion was successful. Numerically found shift from the exact
resonance was in an excellent agreement with the value predicted
by the analytical theory.

In dynamical modeling of the debris disks is improbable that
the dust particle captured to the exterior MMR after random
capture process obtains conditions leading to a permanent capture
with the zero libration amplitude (after averaging over the synodic
period). However, the dust particles can be placed in the vicinity
of such initial conditions (see Fig. \ref{fig:stationarity}).
All captures in the exterior MMRs in the planar CRTBP are unstable in
the sense that the libration amplitude increases in accordance with
results of \citet{gomes95}. Even small libration amplitude will be finally
increased and the capture leads to a close encounter with the planet
(in Fig. \ref{fig:radiation} the $kh$ point crosses the dashed line).
The periodicity discussed in this study belongs to exact
initial conditions that give the zero libration amplitude.
The zero libration amplitude does not increase \citep{gomes95}.

Interesting result is that the existence of periodic motions in
the reference frame orbiting with the planet results from the averaged
theory. The existence of the periodical motions in the MMRs occurring
in the planar CRTBP with radiation demonstrated by the existence
of circular orbits for the 1/1 resonance \citep{points} would be nicely
completed by the existence of the periodical motions in the exterior
resonances. The stationary solution for a given dust particle captured
in a given exterior MMR in the planar CRTBP with radiation
would represent a core that is a periodic motion to which
the eccentricity asymptotically approaches and around
which the libration occurs.

\appendix

\section{Secular time derivatives in rotational symmetry systems}
\label{sec:longitude}

The Gaussian perturbation equations of celestial mechanics are
\citep[cf. e.g.][]{danby,fund}
\begin{align}\label{gauss}
\frac{da}{dt} = {} & \frac{2}{n \sqrt{1 - e^{2}}}
      \left [ a_{\text{R}} e \sin f + a_{\text{T}}
      \left ( 1 + e \cos f \right ) \right ] ~,
\notag \\
\frac{de}{dt} = {} & \frac{\sqrt{1 - e^{2}}}{a n}
      \biggl [ a_{\text{R}} \sin f + a_{\text{T}} \biggl ( \cos f
\notag \\
& + \frac{e + \cos f}{1 + e \cos f} \biggr ) \biggr ] ~,
\notag \\
\frac{d \tilde{\omega}}{dt} = {} & - \frac{\sqrt{1 - e^{2}}}{a n e}
      \biggl ( a_{\text{R}} \cos f
\notag \\
& - a_{\text{T}} \sin f \frac{2 + e \cos f}{1 + e \cos f} \biggr ) ~,
\notag \\
\frac{d \sigma_{\text{b}}}{dt} + t \frac{dn}{dt} = {} & \frac{1 - e^{2}}{a n}
      \biggl [ a_{\text{R}} \left ( \frac{\cos f}{e} -
      \frac{2}{1 + e \cos f} \right )
\notag \\
& - a_{\text{T}} \frac{\sin f}{e}
      \frac{2 + e \cos{f}}{1 + e \cos f} \biggr ] ~,
\end{align}
where $f$ is the true anomaly, $a_{\text{R}}$ is the radial component
of the perturbing acceleration and $a_{\text{T}}$ is the transversal
component of the perturbing acceleration. Let the particle be in some
position in the Keplerian orbit which has the longitude of pericenter
$\tilde{\omega_{1}}$. For the perturbing acceleration of a dust particle
caused by the non-gravitational effects we have in Cartesian coordinates
defined in the orbital plane
\begin{equation}\label{acceleration}
\vec{a}_{\text{P} 1} = ( a_{\text{P} 1 x}, a_{\text{P} 1 y} ) ~.
\end{equation}
Orthogonal radial and transversal unit vectors of the dust particle in
the orbit are
\begin{align}\label{univec}
\vec{e}_{\text{R} 1} &= \left [ \cos \left ( f + \tilde{\omega}_{1} \right ),
      \sin \left ( f + \tilde{\omega}_{1} \right ) \right ] ~,
\notag \\
\vec{e}_{\text{T} 1} &= \left [ - \sin \left ( f + \tilde{\omega}_{1} \right ),
      \cos \left ( f + \tilde{\omega}_{1} \right ) \right ] ~.
\end{align}
We can calculate $a_{\text{R} 1}$ and $a_{\text{T} 1}$ using
\begin{align}\label{scalar}
a_{\text{R} 1} &= \vec{a}_{\text{P} 1} \cdot \vec{e}_{\text{R} 1} =
      a_{\text{P} 1 x} \cos \left ( f + \tilde{\omega}_{1} \right ) +
      a_{\text{P} 1 y} \sin \left ( f + \tilde{\omega}_{1} \right ) ~,
\notag \\
a_{\text{T} 1} &= \vec{a}_{\text{P} 1} \cdot \vec{e}_{\text{T} 1} = -
      a_{\text{P} 1 x} \sin \left ( f + \tilde{\omega}_{1} \right ) +
      a_{\text{P} 1 y} \cos \left ( f + \tilde{\omega}_{1} \right ) ~.
\end{align}
Since the problem has the rotational symmetry it is possible to obtain
the perturbing acceleration in an orbit with different longitude
of pericenter $\tilde{\omega_{2}}$ by a rotation around the origin
of the Cartesian coordinates by the angle $\tilde{\omega_{1}}$ $-$
$\tilde{\omega_{2}}$.
\begin{align}\label{rotation}
a_{\text{P} 2 x} &= a_{\text{P} 1 x} \cos \left ( \tilde{\omega}_{1} -
      \tilde{\omega}_{2} \right ) +
      a_{\text{P} 1 y} \sin \left ( \tilde{\omega}_{1} -
      \tilde{\omega}_{2} \right ) ~,
\notag \\
a_{\text{P} 2 y} &= - a_{\text{P} 1 x} \sin \left ( \tilde{\omega}_{1} -
      \tilde{\omega}_{2} \right ) +
      a_{\text{P} 1 y} \cos \left ( \tilde{\omega}_{1} -
      \tilde{\omega}_{2} \right ) ~.
\end{align}
The radial component of the perturbing acceleration for the particle on
the orbit with the longitude of pericenter $\tilde{\omega}_{2}$ is
\begin{align}\label{Rrotated}
a_{\text{R} 2} = {} & \vec{a}_{\text{P} 2} \cdot \vec{e}_{\text{R} 2}
\notag \\
= {} & \left [ a_{\text{P} 1 x} \cos \left ( \tilde{\omega}_{1} -
      \tilde{\omega}_{2} \right ) +
      a_{\text{P} 1 y} \sin \left ( \tilde{\omega}_{1} -
      \tilde{\omega}_{2} \right ) \right ]
\notag \\
& \times \cos \left ( f + \tilde{\omega}_{2} \right )
\notag \\
& + \left [ - a_{\text{P} 1 x} \sin \left ( \tilde{\omega}_{1} -
      \tilde{\omega}_{2} \right ) +
      a_{\text{P} 1 y} \cos \left ( \tilde{\omega}_{1} -
      \tilde{\omega}_{2} \right ) \right ]
\notag \\
& \times \sin \left ( f + \tilde{\omega}_{2} \right )
\notag \\
= {} & a_{\text{P} 1 x} \cos \left ( f + \tilde{\omega}_{1} \right ) +
      a_{\text{P} 1 y} \sin \left ( f + \tilde{\omega}_{1} \right ) =
      a_{\text{R} 1} ~.
\end{align}
Similarly for the transversal component of the perturbing acceleration
we obtain
\begin{equation}\label{Trotated}
a_{\text{T} 2} = a_{\text{T} 1} ~.
\end{equation}
We come to the conclusion that the radial and transversal components
of the perturbing acceleration are independent of the longitude
of pericenter. Because the longitude of pericenter is not explicitly
present on the right-hand sides of the Gaussian perturbation equations
given by Eqs. (\ref{gauss}), after the averaging process we obtain
the secular expressions independent of the longitude of pericenter.

\section*{Acknowledgments}

The author is grateful to the Nitra Self-governing Region for support.
I would also like to thank the referee for a very detailed and useful report
which helped me to improve the manuscript.

\label{lastpage}

\end{document}